\documentclass[aps,pra,showpacs,twocolumn, superscriptaddress]{revtex4-1}
\usepackage{epsfig,amsmath}
\usepackage{subfigure}
\usepackage{graphicx}
\usepackage{dcolumn}
\usepackage{stmaryrd}
\usepackage{mathrsfs}
\usepackage{pifont}
\usepackage{amsthm}
\usepackage{amssymb}
\usepackage{bm}
\usepackage{latexsym}
\usepackage{hyperref}
\usepackage{color}

\newcommand{\beq}{\begin{equation}}
\newcommand{\eeq}{\end{equation}}
\newcommand{\beqa}{\begin{eqnarray}}
\newcommand{\eeqa}{\end{eqnarray}}

\newcommand{\da}{\dagger}

\begin{document}

\title{\Large{Self-protected quantum simulation and quantum phase estimation in the presence of classical noise} }
\author{Lian-Ao Wu \footnote{lianao.wu@ehu.es}
}
\affiliation{ Department of Physics, University of the Basque Country UPV/EHU, 48080 Bilbao, Spain
IKERBASQUE Basque Foundation for Science, 48013 Bilbao, Spain
EHU Quantum Center, University of the Basque Country UPV/EHU, Leioa, Biscay 48940, Spain}
\date{\today }


\date{\today}
\begin{abstract}
Decoherence is a major challenge in quantum computing. To enable execution of quantum algorithms, it is crucial to eliminate decoherence and noise for instance via dynamic decoupling and quantum error correction protocols based on dynamic zero-noise strategy. As potential alternatives we introduced self-protected quantum algorithms over 15 years ago. Quantum algorithms of this kind, based on the living-with-noise strategy, are now used in the Noisy Intermediate-Scale Quantum regime. Here we introduce self-protected quantum simulations in the presence of weak classical noises. Notably, we prove the equivalence between weak classical noise and noiseless quantum simulations. This equivalence implies that a self-protected quantum simulation does not require any extra overhead in its experimental implementation.  Furthermore, we find that the conventional quantum phase estimation can be upgraded to its corresponding noisy version.


\end{abstract}
\maketitle


{\em Introduction.---} Interest in quantum computing is motivated by a few classes of quantum algorithms that provide speed-up over their classical counterparts.  The fully convincing reasons for this speed-up are still under exploration. However, these algorithms and associated techniques have shown significant benefits to related fields~\cite{Feynman82, Shor97, Grover96}. The best known quantum algorithm is Peter Shor's 1994 ~\cite{Shor94} theoretical demonstration that a quantum computer can solve the hard problem of finding the prime factors of large numbers exponentially faster than all classical counterparts. Since prime factorization is at the center of breaking the ubiquitous RSA-based cryptography,  Shor's factorization scheme immediately attracted great attention, triggering considerable research enthusiasm on quantum computing.  Another class of algorithms, derived from Grover's search algorithms~\cite{Grover96}, can be applied to a wide range of problems where searching for a solution set is an optimal known problem-solving strategy.  A third class is to simulate quantum systems.  Simulation algorithms offer exponential speedups of any known classical algorithms for various quantum systems~\cite{Lloyd96} and are promising for many applications in the physical sciences~\cite{Wu02}. These include molecular, solid-state and nuclear simulations and may not necessarily require fully scalable quantum computing devices.

Ideal quantum computers with complete quantumness are essential to achieve the speedup. However, decoherence and quantum noise plague the possibility of reliable quantum process and lead to loss of quantumness in a short time. This has left us to pin our hope on the use of the quantum error correction initiated by Shor~\cite{Shor96} and subsequent work that extends the method~\cite{Steane98, Knill01}, and helps to scale up quantum computers.   There are also alternative protocols, such as open-loop decoherence-free subspaces (DFSs)~\cite{Zanardi97} and dynamical decoupling (DD)~\cite{Spinecho,Pavel}. These approaches are rooted in a {\em dynamic zero noise strategy} when executing a quantum algorithm. The goal of the strategy is to control and maximally suppress noise to near-zero levels and/or to quarantine the quantum algorithm processing in ``mobile shelters" (DFSs). Scientists believed that the error correction codes, as a benchmark of {\em the dynamic zero noise strategy},  would be easy because physics allows it. Yet in practice, this active closed-loop prevention is extremely difficult, in particular current machines are not even equipped to implement these codes due to for example physical error rate far over threshold and/or limited/poor control~\cite{hype},
The status quo today is that we are in a difficult era, with hundred (s) of noisy qubits/in the so-called Noisy Intermediate-Scale Quantum (NISQ) regime. While the qubit systems we have achieved today are impressive, they have yet to bring us significantly closer to executing any quantum algorithm that anybody cares about. 

While it is unclear that the dynamic zero noise strategy would win the future though ones have great confidence~\cite{hype}, we had thought of the alternative/opposite strategy: the {\em living-with-noise} strategy, 15 years ago. Forward thinking with this strategy, we proposed self-protected quantum algorithms in the year 2007, featured with innate fault-tolerance, self-protection or immunity~\cite{Wu09}. Since then there have been many algorithms developed grounded with the living-with-noise strategy and our self-protection idea, in particular those self-protected algorithms against the specific noise in contemporary noisy intermediate-scale quantum (NISQ) computers~\cite{Barry,Preskill} though named differently.  

Below, under our {\em living-with-noise} strategy, we propose self-protected quantum simulations immune to general weak classical noise. We consider efficiently simulatable ({\em via} either digital simulation or analogy simulation) Hamiltonians and illustrate them for different physical systems. Under dressing transformations~\cite{Wudress}, equivalently these Hamiltonians are embedded in a class of classical noise. By simulating the propagators of these Hamiltonians under classical noise, the spectra of these Hamiltonians can be readout regardless of existence of these classical noises.  To read out spectra, one can either measure voltage signal induced by precessing magnetic moment as done in an NMR quantum computer, or use quantum phase estimation (QPE).  We prove that the conventional quantum phase estimation (QPE) can be generalized to an upgraded version in the presence of classical noise. Note that our concerned Hamiltonians should not belong to the class of exactly solvable models~\cite{Wutalk} which are ready to be solved efficiently with a classical computer.  Significantly, the equivalence between weak classical noise and noiseless quantum simulations is based on mathematical proof and does not require any additional experimental operation.


{\em Classical noise.---} Decoherence is modelled by the first-order interactions between a system and its environment. Methods to combat decoherence, for instance the QECC, are therefore designed based on this linear decoherence model with the first-order interaction $H_{I}=\vec{\sigma} \cdot \vec{B}$ between a qubit and its environment. The total Hamiltonian of a system and its environment is conventionally written as, 
\begin{equation}
H_{total}=H+H_B+H_{I}, 
\end{equation}
where $H (H_B)$ is the system (environment) Hamiltonian. The interaction takes the form of $H_I = \sum_jA_j B_j$, where $A_j$'s are Hermitian operators in the Hilbert space of the system, for example $\vec{A}=\vec{\sigma}$ for a single qubit, and $B_j$'s are environmental operators. When $H_I$ can be semi-classically approximated by $H_I = \sum_jA_j\langle B_j\rangle$, where $\langle B_j\rangle$'s are now $c$-numbers determined by the random states of the time-independent environment, this yields a stochastic Hamiltonian acting on the system, $H_a=H+ H_I$, where the system bare Hamiltonian $H$ is time-independent. The parameter set $a=(a_1,a_2,...)$ in $H_a$ represents random parameters, where $a_j$ are in correspondence with $\langle B_j\rangle$~\cite{Jing17}. The reliability of the semi-classical approximation has been studied. For example Ref. ~\cite{Franco19} shows that high-temperature quantum decoherence, that obviously alleviates experimental constraints regarding the selection of related simulators or setups and provides more feasibility in the experimental process, can be treated by classical coloured Gaussian noise.

Determined by $H_a$ corresponding to a particular realization (or channel) $a$, the system evolution  is unitary, $\rho^a \mapsto W(a)\rho^{0}W(a)^\da$. Here $a$ corresponds to the classical stochastic environmental parameters and $W_{t\geq0}(a)$ is defined as a propagator $e^{-iH_at}$ with $\hbar\equiv1$.  The configuration can be conventionally assumed to be a probability distribution $p_a$. The state of the system at time $t$ is the average over all possible unitary evolutions 
\begin{equation}\label{Kraus}
\rho(t)=\sum_a p_ae^{-iH_at}\rho^{0} e^{iH_at},
\end{equation}
with $\sum_a p_a =1$. This is a general expression, regardless of the details of $H$. 

From now on we will focus on a wide variety of classical noise that becomes {\em general} in the linear coupling $H_I = \sum_jA_j\langle B_j\rangle$ when the noise is weak enough, as shown in the examples below,
\begin{equation}\label{main}
H_a=V_aHV^{\dagger}_a,
\end{equation}
introduced by the dressing transformations $V_a$. The noise features that the noiseless $H$ and the noisy $H_a$ have the same eigenvalues. It is clear that $H_a$ can be always written as $H_a=H+H_I$, which can be seen from the following physical realizations.


{\em Physical realizations.---}Classical noiseWe first consider the eigenproblem of a class of many-spin Hamiltonians represented by Pauli $\sigma^x$ and $\sigma^z$ matrices,
\begin{equation}
 H = B \sum_i \sigma^x_i + h( { \sigma^z_ i } ),
 \end{equation}
where $h({\sigma^z_i})$ contains the $\sigma^z_i$ component of the $i$-th qubit. When the parameters $B$ is time-dependent, it could represent an instantaneous Hamiltonians in adiabatic quantum computing (AQC). The solution of a hard problem is encoded within $h$. For example, Grover's search problem is realized with
$h( { \sigma^z_ i } ) = I-|B\rangle \langle B|$,  where $|B\rangle$ is the index state, and $|B\rangle \langle B|$ is a function of $\sigma^z_i$.
Alternatively, in the D-Wave system the Hamiltonian (3) is written as~\cite{Dwave}
\begin{equation}
h( { \sigma^z_ i } )=\sum_i h_i \sigma^z_i +\sum_{ij} J_{ij} \sigma^z_i \sigma^z_j,
 \end{equation}
which contains the solution of an optimization problem. 
We here focus specifically on energy spectra, so $B$ and $h$ are set to be time-independent. For these situations, the transformation $V_a=\prod_i e^{-ia_i \sigma^z_i}$ introduces noisy Hamiltonians $H_a = H+B\sum_i [(\cos a_i-1) \sigma^x_i+ \sin a_i \sigma^y_i ]$, which becomes 
\begin{equation}
 H_a = H+B\sum_i  a_i \sigma^y_i 
 \end{equation}
for weak enough random parameters $a_i$, where $a=(a_1,a_2,...)$ is a set of random parameters characterizing a particular realization of the system evolution.  Taking $B a_i =\langle \lambda_i ( b^{\dagger}_i+b_i ) \rangle$, the set $a$ thus comes from the classicalized  famous Spin-Boson model, where $b_i$ are bath Boson operators. The famous spin-boson model has been demonstrated to be a perfect characterization for spin and its physical bath, and therefore it is reasonable to believe its classialization well fits to the behind physics and is {\em general.}  


Second, an interacting many-body problem can be given by a generic Hamiltonian
\begin{equation}\label{q}
 H =\sum_i \frac{p_i^2}{2m_i}+V_i(x_i)+\sum_{ij}V_{ij}(|x_i-x_j|)
 \end{equation}
The first (second) term is the kinetic energy (potential). If $V_i$ has a minimum at $X_i$, we can Taylor expand it about the minimum and take an approximation $V_i(x_i)\approx V_i(X_i)+\frac{k_i^2}{2}(x_i-X_i)^2$, where $k>0$ is the second derivative of $V_i$ at $x_i=X_i$. Without loss of generality, we can always choose $V_i(X_i)=0$. The last term is the interaction between particles $i$ and $j$. Similarly, we can Taylor expand $V_{ij}(|x_i-x_j|)\approx \sum_{ij} \frac{k_{ij}}{2}(x_i-x_j)^2$. In principle, decoherence supposedly comes from the system-bath coupling in the form of $\sum_i A_i p_i/m_i$, for instance $\vec{A}\cdot\vec{p}/m$ where 
$\vec{A}$ is the vector potential of the bath particles such as phonons. The field $A$ can be classicalized and expressed by $a=(a_1,a_2,...)$. By the dressing transformation $V_a= \prod_j e^{-ia_j m_jx_j}$, the noisy Hamiltonian can be given,
\begin{equation}
 H_a=H+\sum_j a_j p_j,
\end{equation}
where we ignore the constant in the Hamiltonian. The term $\sum_j a_j p_j$ corresponds to a classicalized vector potential.  

Another independent linear term can be introduced from the transformation  
$V_a=  \prod_j e^{-ia_j m_jx_j} \prod_j e^{-ib_j p_j/k_j}$,  the noisy Hamiltonian thus is
\begin{equation}\label{general}
 H_a=H+\sum_j (a_j p_j+a'_j x_j)
\end{equation}
where, after tedious derivations,  $a'_j=b_j +\sum_l k_{jl}(b_j-b_l)/k_l)$ and $a=(a_1,a_2,...,a'_1,a'_2,...)$. The noisy Hamiltonian is generic in the sense that all the linear couplings between the system and its bath are considered and independent when bath classicalization is taken. 


As a third example, we would simulate the famous spin boson model,
\begin{equation}
 H = B \sigma^x+ \sigma^z \sum_{\alpha=1} \lambda_k (b_\alpha+b^{\dagger}_\alpha)+\sum_{\alpha=1} \omega_\alpha b^{\dagger}_{\alpha} b_{\alpha},
\end{equation}
under classical noise, where $b_{\alpha} (b^{\dagger}_{\alpha})$ the annihilation (creation) operators for a boson in the single-particle state $\alpha$. 
Noise can be introduced by the dressing transformation $V_a= e^{-ia_0\sigma^z}\prod\limits_{\alpha=1} e^{a^*_{\alpha}b_{\alpha}-a_{\alpha}b^{\dagger}_{\alpha}}$ and the noisy Hamiltonian reads 
$H_a = H+\vec{\sigma} \cdot \vec{B}+\sum_{\alpha}(\omega_\alpha a_\alpha b^{\dagger}_\alpha +h.c.)$
where $ \vec{B}=(B(\cos a_0-1),  B\sin a_0, \sum_{\alpha}\lambda_{\alpha}(a_\alpha+a^*_\alpha ))$ and the random parameter set $a=(a_0,a_1,....)$.  All system variables $\vec{\sigma}$ and $b_{\alpha}, b^{\dagger}_{\alpha}$ are linearly coupled to their corresponding environments. 
However, for weak random parameters
\begin{align}
H_a = H+\sigma_yBa_0+\sigma_zB \sum\lambda_{\alpha}(a_\alpha+a^*_\alpha )     \nonumber \\
+\sum_{\alpha=1}(\omega_\alpha a_\alpha b^{\dagger}_\alpha +h.c.). 
\end{align}
It becomes {\em general} as discussed in the first example. The model can of course be generalized to multi-spin-boson models.

{\color{red}{In general, for an arbitrary $V_a=e^{-i\epsilon P_a}$ with weak strength $\epsilon$, the first order is $H_a\approx H+i\epsilon [H,P_a] $. For a Hamiltonian $H$, as long as the first-order term is incorporated into the linear coupling $ i\epsilon [H,P_a] \sim  \sum_jA_j\langle B_j\rangle$,  $H$ can belong to our class of self-protected quantum simulations, as shown in the above three examples.}}

In the fourth example, we focus on one of the greatest models in the last century is the BCS-type Hamiltonian for superconductivity~\cite{WuJPA},
\begin{equation}
H=\sum_{k,\sigma =\downarrow ,\uparrow }\epsilon _{k}n_{k\sigma }+\sum_{k,k^{\prime }}V_{k,k^{\prime }}\eta _{0}^{\dagger }(k^{\prime
})\eta _{0}(k),
\end{equation}
where $\eta _{0}(k)$ is expressed in terms of the pair operators 
\begin{equation}
\eta _{q}(k)=c_{k\uparrow }c_{q-k\downarrow }\text{ and }\eta _{q}^{\dagger
}(k)=c_{q-k\downarrow }^{\dagger }c_{k\uparrow }^{\dagger },  \label{qpair}
\end{equation}%
when $q=0$. The vectors or modes $k=$ $(k_{x},k_{y},k_{z})$ with $%
k_{x,y,z}=2\pi l/L$, where $l=0,1,...,L-1$. Vectors $q$ have the same modes.
The model is reduced to the standard BCS model when $V_{k,k^{\prime}}= G$ a constant. 
Noise could bring the BCS pairs deviating from the exact $q=0$, i.e.,  $\eta _{q}(k)$. As shown in our paper~\cite{WuBCS}, the conventional pairing model can be in principle simulated by a quantum computer and have been implemented experimentally~\cite{NMR}. Noise is introduced by the unitary transformation, 
\begin{equation}
V_q=\exp (\frac{\pi }{2}[\sum_{k}g(k)(c_{q-k%
\downarrow }^{\dagger }c_{q^{\prime }-k\downarrow }-c_{q^{\prime
}-k\downarrow }^{\dagger }c_{q-k\downarrow })])
\end{equation}
where $a=(q_1,q_2,...)$ can be either discrete or continuous. 
The corresponding Hamiltonian of noise channel $q$ is $H_q=V_qHV^{\dagger}_q$, 
\begin{equation}
H_q=\sum_{k,\sigma =\downarrow ,\uparrow }\epsilon _{k}n_{k\sigma }+\sum_{q,k,k^{\prime }}V_{k,k^{\prime }}\eta _{q}^{\dagger }(k^{\prime
})\eta _{q}(k)
\label{general}
\end{equation}
Under the transformations $V_q$, the
single-particle energies are likewise invariant,%
\[
V_{q}\sum_{k,\sigma =\downarrow ,\uparrow }\epsilon
_{k}c_{k\sigma }^{\dagger }c_{k\sigma }V_{q}=\sum_{k,\sigma
=\downarrow ,\uparrow }\epsilon _{k\sigma }(q)c_{k\sigma }^{\dagger
}c_{k\sigma } , \label{main1} 
\]%
where $\epsilon _{k\sigma }(q)=\epsilon _{k}\delta _{\sigma \uparrow
}+\epsilon _{k-q}\delta _{\sigma \downarrow } $. It remains diagonal but
becomes spin-dependent. This implies that the $q$-pairing Hamiltonian $H_{q}$
is equivalent to a BCS-type Hamiltonian with a spin-dependent single-particle
energy. 

The model essentially means classical fluctuation around $q=0$. It is worth pointing out that intuitively the fourth example does not seem to fit the conventional decoherence model, where the system and its environment are linearly coupled. Yet it shows our modelling is also applicable to decoherence other than the conventional linear-coupling model.


{\em Readouts.---} As for readout, one can measure the voltage signal induced by precessing magnetic moment. The principal output is the free induction decay signal,
\begin{equation}
\begin{split}
V(t)=V_0 \sum_a p_a \text{Tr} [e^{-iH_a t}\rho^{0}e^{iH_a t}O ]\\
=V_0 \sum_a  p_a \text{Tr} [e^{-iH t}\rho^{a}e^{iH t}O^{a}] \\
=V_0 \sum_a p_a \rho^{a}_{uv}O^{a}_{vu}e^{-i\omega_{uv}t}
\end{split}
\end{equation}
where $\omega_{uv}=E_u -E_v$ correspond to spectra of the original Hamiltonian $H$ and $\rho^{a}_{uv}=\sum_{kl}\rho^{0}_{kl} \langle u | V_a | k \rangle \langle l | V^{\dagger}_a |v\rangle $, so are  $O^{a}_{uv}$. This signal could be measured using quantum state tomography for some period of time, and then the results are Fourier-transformed into a plot. The operator $O$ is often $O=\sigma^x_k+i\sigma^x_k$ for a qubit system. As a shortcoming, the method does not give the absolute values of system energies. The advantage is that energy spectra can be given directly from the  simulated system without introducing additional degrees of freedoms. The quantum simulation and readout do not rely on a quantum computer.

Now we come to generalize the conventional quantum phase estimation (QPE) with a mixed initial state. Here we term it as {\em generalized} QPE. The convenient QPE procedure uses two registers: the first register with $t$ qubits begins in the state $|{\bf 0} \rangle$ and the second register, which is not necessary a qubit system,  starts with a superposition of energy eigenstates, 
$| I \rangle = |{\bf 0} \rangle \sum_u C_u | u\rangle $
where  $| u\rangle $ is an eigenstate of $H$. It is easy to show $H_a V_a |u\rangle =V_aHV_a^{\dagger} V_a |u\rangle =E_u V_a  |u\rangle$. For the channel $a$ the initial state is rewritten as  
\begin{equation}
| I \rangle ^{a} = |{\bf 0} \rangle \sum_u C_u^{a} V_a | u\rangle,
\end{equation}
for $H_a$, where the coefficients for the channel $a$ is 
\begin{equation}
C_u^{a}=\sum_{uv}(V_a^{\dagger})_{uv}C_v, 
\end{equation}
which is a linear combination of the noiseless $C_v$ via the dressing transformation $V_a$. Each channel $a$ including the noiseless channel itself therefore undergoes the same procedure as done in the conventional QPE.

For the noiseless channel, as convention by defining $E_ut=- {\pi j_u}/{2^{n-1}}$ for $e^{-iE_ut}$ located in the first register with $n$ qubits and applying black box carried by the second register ( {\em i. e.} a series of controlled-$U$ gates), the whole system becomes 
\begin{equation}
\frac{1}{2^{n/2}}\sum_u C_u \sum_{k=0}^{2^n-1}e^{-i E_utk} |k\rangle | u\rangle 
\end{equation}
The result of running the convenient QPE will be to give as the output state 
\begin{equation}
| O^0 \rangle = \sum_{u}|j_u\rangle C_u  | u\rangle 
\end{equation}
for the noiseless channel. And by measuring $j_u$, the energy $E_ut$ are given. The corresponding density matrix is 
$ \rho^{0} =| O^{0} \rangle   \langle O^{0} | $. 

Similarly, the density matrix for channel $a$ is $ \rho^{a} =| O^{a} \rangle   \langle O^{a} | $ and the final output is
\begin{equation}
\sum_{uv,a} p_aC^{a}_u C_v^{*a} | u\rangle \langle v | |j_u\rangle \langle j_v | 
\end{equation}
Final measurement of this mixed state will give one of ${j_u}$'s. Note that $a$ in the sum $\sum_{uv,a} p_aC^{a}_u C_v^{*a}$ is a dummy index and does not contribute the probability to obtain a given $j_u$.

The {\em generalized} QPE is a standard protocol in obtaining spectra of classical-noisy Hamiltonians, in particular as a standard tool for the generic Hamiltonian~(\ref{general}).


{\em Conclusion.---} A universal fault-tolerant quantum computer can solve efficiently problems such as integer factorization but requires millions of qubits with physical error rates below a certain threshold and long coherence times. Experimental realizations of such devices are still far from reach, yet fortunately the so-called noisy intermediate-scale quantum (NISQ) devices already exist~\cite{Preskill}. These devices are composed of hundreds of noisy qubits, i.e. qubits that are not error-corrected. Therefore it will be significant if ones can find innate fault-tolerant, self-protected or immune quantum algorithms as many as possible, in particular those compatible with existent NISQ hardwires. We should strive to find self-protected versions of known quantum algorithms such as the Grover's algorithm, with the hope that someday the living-with-noise strategy may potentially have a significant impact .

We introduced self-protected quantum algorithms, based on the {\em living-with-noise} strategy,  for factoring multi variable polynomials and obtaining the expectation values of an observable in the presence of a wide variety of noises or errors 15 years ago. The simulation algorithms we introduce here based on the same {\em living-with-noise} strategy are clearly polynomial while the best known classical counterparts of the simulation algorithms are surely exponential. Some of simulated Hamiltonians under classical noise are even fully generic assuming that there are only linear couplings between the system and its bath. The self-protected quantum simulation, along with the previously developed self-protected algorithms, presents a promising alternative to the protocols based on the dynamic zero noise strategy.

    The work is supported by the Spanish Grant No. PID2021- 126273NB- I00 funded by MCIN/AEI/10.13039/501100011033, and the Basque Government through Grant No. IT1470-22.

\end{document}